# The covalent radii derived from the first-principle data

Tymofii Yu. Nikolaienko[1*], Valerii S. Chuiko[1], Leonid A. Bulavin[1]

**Abstract**

We present a collection of covalent radii for the elements H, B, C, N, O, F, Si, P, S, Cl, Ge, As, Se, Br, derived from the recently introduced systematic non-empirical dataset of the covalent bond lengths. As the underlying bond lengths dataset was built basing on 26050 molecular geometries and bond graphs data obtained from quantum-chemical calculations with almost no input of empirical data, the presented covalent radii can be referred to as the first-principle ones. The obtained first-principle covalent radii are in good agreement with their empirical counterparts available in the literature. The distributions of deviations between the sum of proposed radii and the true bond lengths are analyzed in details, with emphasis put on the role of the electron delocalization on the bond lengths. Corrections to the additive covalent radii model due to the difference in the elements electronegativities are derived for both Pauling and Mulliken electronegativity scales.

**Introduction**

Among the different descriptors introduced so far[1–7] to facilitate quantitative characterization of the covalent bonding concept, the atomic covalent radii stand out as one of the simplest, yet widely used, parameters enabling prediction of the bond length. The additive approximation[8–16]

$$l_{AB} = r^{cov}_A + r^{cov}_B, \qquad (1)$$

for the length $l_{AB}$ of the covalent bond formed by the two atoms $A$ and $B$, with $r^{cov}_A$ and $r^{cov}_B$ standing for the covalent radii of the chemical elements, was first introduced nearly a century ago[17,15,16] as a result of generalization of the available experimental data on molecular structure. As quantitative properties of the chemical elements, the covalent radii not only do have methodological value in demonstrating the periodical properties of the elements[18], but are also of practical importance due to widespread use of the approximation (1) in molecular modeling and crystallographic software for distinguishing between bonded and non-bonded atomic pairs[19] and subsequent resolving of molecular structures[14,20], as well as for generating 3D atomic coordinates based on the molecular 'line formula'.

Extensions of the additive model improving the accuracy of the approximation (1) exist[21,22], e.g., accounting for the bond shortening due to the elements electronegativity difference $\chi_A - \chi_B$ as

$$l_{AB} = r^{cov}_A + r^{cov}_B - \beta \cdot |\chi_A - \chi_B|. \qquad (2)$$

Yet, even the simplest approximation reveals (1) an important feature of these models: they all rely on some dataset of the covalent bond lengths in order to derive the values of the elements radii or other empirical descriptors assigned to the chemical elements by these models. Not only must such dataset contain enough data to find all the parameters for all chemical elements and to avoid the risk of an overfitting, but the bonds present in the dataset must be clearly distinguished by the bond multiplicity (e.g., single, double, triple bonds), as the resulting covalent radii for these bond differ substantially. Although determination of the integer-valued bond multiplicity is straightforward when the structural formula of the molecule is known, it can be non-trivial when atomic coordinates is the only available information about the molecules contained in the dataset. The mentioned requirements have so far limited the available datasets to those created by careful manual selection of molecules and curated assignment of multiplicities to their bonds.

Alternative approach has been proposed recently for non-empirical determination of the bonding graph and integer-valued bond multiplicities[23] by automated analysis of molecular electronic structure in the basis of orbitals found by the property-optimized orbitals localization procedure. This approach was used to identify the covalent bonds in a collection of 26050 molecular structures, containing from 2 to 12 atoms and optimized by one of the three different DFT methods (B3LYP/6-31G*, PBEh-3c, B97-3c). In this way, a systematic dataset was created[23] containing the lengths of 213404 single, 38732 double and 5031 triple bonds formed by H, B, C, N, O, F, Si, P, S, Cl, Ge, As, Se, Br elements.

---

[1] Faculty of Physics of Taras Shevchenko National University of Kyiv; 64/13, Volodymyrska Str., 01601 Kyiv, Ukraine; Correspondence to: tim_mail@ukr.net

Availability of this extensive dataset allowed revealing some features in statistical distributions of the covalent bond lengths which have been previously overlooked in smaller bond lengths collections. Particularly, it has been found that the distributions for certain bonds have non-trivial (sometimes, multi-peak) shapes which differ from that of a single-peak normal distribution[23]. This fact has important consequences for determining the covalent radii because the estimates based on 'least-squares' (LSE) approach (typically used to determine the radii) and 'maximum likelyhood' (MLE) criteria become different as soon as the underlying data possesses non-Gaussian distribution. On the other hand, it has been found that by sieving out the conjugated bonds (in a sense defined in ref.[23]) it is possible to obtain the subset of the bond lengths collection in which the distributions of the bond lengths are much closer to the normal one.

In this work we use the recently proposed dataset[23] of covalent bond lengths derived from the first-principle data to obtain the 'first-principle' covalent radii and to investigate the influence of bond lengths statistics on their ultimate values. Specifically, the radii derived for the subset of non-conjugated bonds are compared with the radii derived from entire set of the bonds lengths.

**Computation details**

The usual way of determining the atomic covalent radii is by fitting (1) to the set of bond lengths[10–13], i.e., by minimizing the bond length approximation error

$$\varepsilon\left(\left\{r^{cov}_X\right\}\right) = \sum_{\{A \geq B\}} \sum_{i \in AB} \left(r^{cov}_A + r^{cov}_B - l_i\right)^2, \qquad (3)$$

where $\{A \geq B\}$ denotes all possible pairs of the chemical elements, and '$i \in AB$' indicates that the summation index $i$ enumerates the bonds of a given multiplicity (e.g., single) formed by the elements $A$ and $B$.

It is worth adding that minimization of error function (3) is usually performed separately for sets of the single, double and triple bonds[10–13] thereby accounting for systematic bond shortening with increase in the bond multiplicity and leading to up to three different values of covalent radius per chemical element. Importantly, the bond length dataset reported in ref.[23] contains the required integer-valued bond multiplicities, determined by the number of BD CLPOs.

The error function (3) can be rewritten in vector-matrix form as

$$\varepsilon\left(\left\{r^{cov}_X\right\}\right) = \mathbf{r}^T \mathbf{A} \mathbf{r} - 2\mathbf{r}^T \mathbf{b} + \varepsilon_0, \qquad (4)$$

where $\mathbf{r}$ is the column-vector containing the radii to be determined, $\mathbf{A}$ is the square matrix with the elements determined by the number of bonds formed by the particular pair of chemical elements, $\mathbf{b}$ is the column vector whose elements are determined by the sums of lengths of the bond in which the particular elements is involved, and $\varepsilon_0$ is the constant, independent of $\mathbf{r}$. In the Eq. (4) the lengths of the bonds formed by particular pair of chemical elements contribute only to one of the elements of $\mathbf{b}$ and to the constant term $\varepsilon_0$.

Minimization of either (3) or (4) is straightforward and leads to the system of linear equations, allowing for a unique solution

$$\mathbf{r} = \mathbf{A}^{-1}\mathbf{b}.$$

Since the elements of $\mathbf{b}$ involve only the sums of the lengths of particular bond and the elements of $\mathbf{A}$ do not depend on the lengths, it can be concluded that the only structural information on which the ultimate covalent radii depend is the sample-averaged bond lengths $\langle l^{AB} \rangle = \frac{1}{n_{AB}} \sum_{i \in AB} l_i$. The total number of bonds of particular type in the sample on the other hand determines only relative error weightings for these sample-averaged lengths approximation with the ultimate radii.

All computations have been performed using a Python program developed as a part of the present study.

**Results and discussion**

By fitting the lengths of single, double and triple bonds obtained by the proposed method according to (1)–(4) we obtained the 'first-principle' covalent radii presented in Tables 1 and 2. These radii are in very good agreement with those found from experimental bond lengths (typically requiring considerable amount of human-curated data processing), as illustrated by Fig. 1.

Table 1. Covalent radii (in Å) of the chemical elements derived for single, double and triple bonds from the dataset of non-conjugated bonds in the molecules optimized at B97-3c level.

| Chemical element | Single bonds | | Double bonds | | Triple bonds | |
|---|---|---|---|---|---|---|
| | Without electronegativity correction | With electronegativity correction[a] | Without electronegativity correction | With electronegativity correction[a] | Without electronegativity correction | With electronegativity correction[a] |
| H | 0.326 | 0.363 | – | – | – | – |
| B | 0.834 | 0.853 | 0.817 | 0.858 | 0.655 | 0.667 |
| C | 0.757 | 0.759 | 0.659 | 0.663 | 0.600 | 0.603 |
| N | 0.702 | 0.735 | 0.616 | 0.630 | 0.564 | 0.568 |
| O | 0.656 | 0.726 | 0.543 | 0.593 | 0.557 | 0.572 |
| F | 0.603 | 0.742 | 0.510 | 0.601 | – | – |
| SI | 1.090 | 1.129 | 1.050 | 1.079 | 0.979 | 0.996 |
| P | 1.050 | 1.070 | 0.946 | 0.976 | 0.921 | 0.930 |
| S | 1.040 | 1.045 | 0.955 | 0.955 | 0.959 | 0.957 |
| CL | 1.048 | 1.105 | 0.968 | 1.001 | – | – |
| GE | 1.169 | 1.195 | – | – | – | – |
| AS | 1.161 | 1.199 | 1.071 | 1.107 | 1.037 | 1.046 |
| SE | 1.159 | 1.166 | 1.064 | 1.075 | – | – |
| BR | 1.215 | 1.249 | 1.161 | 1.184 | – | – |
| β | – | -0.102 | – | -0.068 | – | -0.020 |

[a] Obtained using Eq. (2) for the fitting, with the values of Pauling electronegativities as reported by Allred in ref.[24].

Table 2. Covalent radii (in Å) of the chemical elements derived for single, double and triple bonds from the whole sample of bonds in the molecules optimized at B97-3c level.

| Chemical element | Single bonds | Double bonds | Triple bonds |
|---|---|---|---|
| N | 0.681 | 0.626 | 0.563 |
| GE | 1.168 | - | - |
| NA | 1.594 | - | - |
| AS | 1.169 | 1.078 | 1.039 |
| SE | 1.157 | 1.065 | - |
| H | 0.341 | - | - |
| S | 1.033 | 0.957 | 0.957 |
| CL | 1.038 | 0.952 | - |
| BR | 1.187 | 1.119 | - |
| LI | 1.390 | - | - |
| AL | 1.109 | - | - |
| B | 0.790 | 0.819 | 0.657 |
| O | 0.644 | 0.537 | 0.554 |
| SI | 1.088 | 1.045 | 0.977 |
| P | 1.050 | 0.956 | 0.923 |
| C | 0.741 | 0.671 | 0.603 |
| F | 0.612 | 0.510 | - |

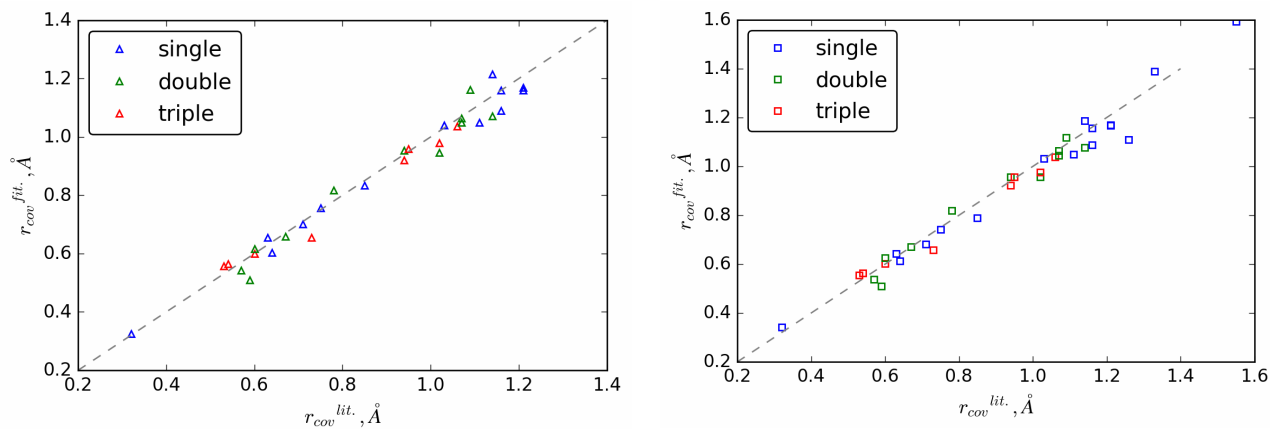

Fig. 1. Comparison of the covalent radii obtained from the samples of non-conjugated bonds (*left*) and from the entire bond length dataset (*right*) with the radii available in the literature[10–13]

Availability of large bond lengths datasets obtained by the presented method allows performing further analysis of errors in approximating the covalent bond length with the sum of the covalent radii. It is worth noting that the total distribution of the fit errors is formed as a superposition of distributions of the bond lengths for different pairs of atoms, each being shifted by a constant equal to the sum of the covalent radii. Consequently, the distribution of fitting errors can only be expected to be of a nearly-Gaussian shape when the distributions for the lengths of each individual bond type all have the same shape as well.

Inspection of the obtained distributions of errors presented in Fig. 2 confirms that this expectation in that the distributions of errors obtained from the sub-sample of non-conjugated bonds have their shapes much closer to single-peak normal distribution as compared to the distributions obtained from the entire collection of bond lengths.

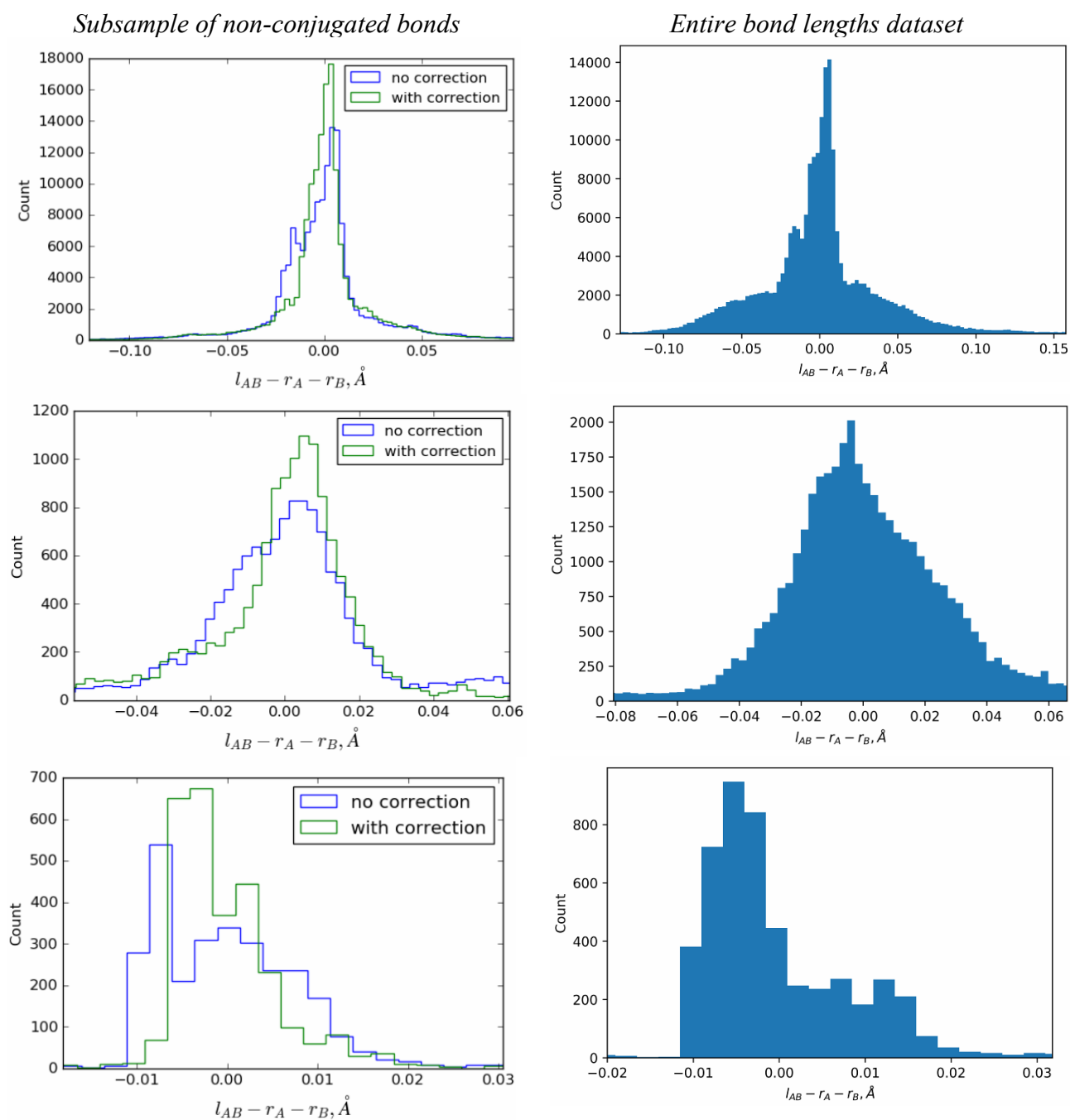

Fig. 2. The distributions of errors in approximating the lengths of covalent bonds with additive covalent radii obtained in the present study from molecular geometries optimized at B97-3c level; from top to bottom: single, double, triple bonds.

In particular, the distributions for the errors in approximating the single and double bonds become more narrow when the initial dataset is limited to non-conjugated bonds only. More importantly, the shape of the error distribution for the fit based on the subset of non-conjugated single bonds is much closer to the single-Gaussian one. Inclusion of bond length shortening correction related to the difference in the bonded elements electronegativities further reduces the widths of the distributions, whereas the ovaral change of their shape is minor.

**Conclusions**

It has been shown that with the obtained dataset of covalent bond lengths resulting from the first-principle calculations it is possible to find non-empirical covalent radii for the chemical element, as well as to derive the corrections to the covalent radii model accounting for the differences in the elements electronegativities. The obtained radii have been found to be in a good agreement with the radii obtained from experimental structural data subjected to human-curated processing.